\documentstyle[12pt]{article}

\begin{document}
\sloppy

\title{Equilibrium Dynamics of Microemulsion and Sponge Phases}
\author{G. Gompper and M. Hennes \\
Sektion Physik der Ludwig-Maximilians-Universit\"at M\"unchen \\
Theresienstr. 37, 80333 M\"unchen, Germany}

\maketitle

\begin{abstract}
The dynamic structure factor $G({\bf k},\omega)$ is studied in a
time-dependent Ginzburg-Landau model for microemulsion and sponge
phases in thermal equilibrium by field-theoretic perturbation methods.
In bulk contrast, we find that for sufficiently small viscosity
$\eta$, the structure factor develops a peak at non-zero frequency
$\omega$, for fixed wavenumber $k$ with $k_0 < k {< \atop \sim} q$.
Here, $2\pi/q$ is the typical domain size of oil- and water-regions in
a microemulsion, and $k_0 \sim \eta q^2$. This implies that the
intermediate scattering function, $G({\bf k}, t)$, {\it oscillates} in
time. We give a simple explanation, based on the Navier-Stokes
equation, for these temporal oscillations by considering the flow
through a tube of radius $R \simeq \pi/q$, with a radius-dependent
tension.
\end{abstract}

\begin{center}
\end{center}
\newpage

\section{Introduction}

\par The behavior of self-assembling amphiphilic systems has attracted
much attention recently \cite{books,GSreview}. The unique properties
of these systems can easily traced back to the ability of the
amphiphile to reduce the interfacial tension between two normally
inmiscible fluids like oil and water by several orders of magnitude.
One of the most interesting phases of ternary amphiphilic systems is
the microemulsion, a homogeneous, isotropic mixture of all three
components. It is now well established that microemulsions with
medium- or long-chain amphiphiles consist of coherent oil- and
water-regions, which are separated by amphiphilic monolayers.
Similarly, the sponge phase in aqueous amphiphilic solutions consists
of distinct water-regions, which are separated by an amphiphilic
bilayer \cite{Porte,sponge}. Thus, the static behavior of these
systems is now relatively well understood. Much less effort has been
made so far, however, to study the dynamical behavior of these
systems. Most of these studies have considered non-equilibrium
situations, typically spinodal phase separation
\cite{Chowdhury,SchBloss}, spontaneous emulsification
\cite{GraBaCates}, oil solubilization in aqueous surfactant solutions
\cite{Shell}, and amphiphile aggregation at an oil/water interface
\cite{Stauffer}.

\par We want to investigate here the dynamic structure factor $G({\bf
k}, \omega)$, both with bulk and film contrast, of microemulsions in
{\it thermal equilibrium}.
It is well known from many studies near critical points that systems
belonging to the same static universality class may show completely
different dynamical behavior \cite{HoHa,Mazenko}. It is thus essential
to clarify which variables have to be included in a dynamic model of
microemulsion and sponge phases. A microemulsion, for example, has
two conserved densities, the amphiphile concentration $\psi({\bf r},
t)$, and the difference of oil- and water-concentration $\Phi({\bf r},
t)$. A sponge phase, on the other hand, may also have two conserved
densities, if the amphiphilic bilayers are almost inpenetrable to
water, or only one, the amphiphile concentration, if they are not
\cite{sponge_dyn}. Nevertheless, the amphiphile concentration $\psi$
may not have to be considered explicitly, if its relaxation is fast
compared to that of the order parameter $\Phi$. This is the case we
want to study in this paper. The opposite limit, where the amphiphile
dynamics is slow compared to the order parameter dynamics, is more
complicated and will be considered elsewhere.
We will show below that it is essential to include the hydrodynamic
variables in a model for microemulsion and sponge phases. In general,
these are the pressure $p({\bf r}, t)$, and the longitudinal and
transverse components of the momentum density, ${\bf j}_L({\bf r}, t)$
and ${\bf j}_T({\bf r}, t)$. It has been shown that in the vicinity of
a critical point, $p$ and ${\bf j}_L$ are irrelevant variables, and
thus can be omitted in the model \cite{HoHa}. We will assume that the
same is the case here. However, $p$ and ${\bf j}_L$ are essential for
studies of sound propagation in these systems
\cite{Kawasaki,KrollRuh,GomHen}.

\section{Ginzburg-Landau Model}

\par The time-dependent Ginzburg-Landau model we consider is a
generalization of the Siggia, Halperin, and Hohenberg model H
\cite{SiHaHo,HoHa}, consistent with linear hydrodynamics. Our
hydrodynamic variables are the local order parameter $\Phi({\bf
r},t)$, which is proportional to the local concentration difference
between oil and water, and the transverse component of the momentum
density ${\bf j}_T({\bf r},t)$. In this case, the stochastic equations
of motion can be written as
\begin{eqnarray} \label{Langevin1}
\frac{\partial \Phi}{\partial t} &=&
  \Gamma_\Phi \nabla^2 \frac{\delta {\cal F}}{\delta \Phi}
+ g_0 \nabla \left(\Phi\frac{\delta {\cal F}}{\delta {\bf j}_T} \right)
+ \zeta_\Phi \\
\label{Langevin2}
\frac{\partial {\bf j}_T}{\partial t} &=&
 \Gamma_T \nabla^2 \frac{\delta {\cal F}}{\delta {\bf j}_T}
+ g_0 \left(\Phi\nabla\frac{\delta {\cal F}}{\delta \Phi} \right)_T
+ \mbox{\boldmath $\zeta$}_T \ ,
\end{eqnarray}
where the Langevin forces $\zeta_i$ have the usual correlations
\cite{SiHaHo},
\begin{eqnarray} \label{noise_phi}
< \zeta_\Phi({\bf r},t) \zeta_\Phi({\bf r}',t') > &=&
   -2 \Gamma_\Phi \nabla^2 \delta({\bf r-r'}) \delta(t-t') \\
\label{noise_T}
< \zeta_{T,\alpha}({\bf r},t) \zeta_{T,\beta}({\bf r}',t') > &=&
   -2 \Gamma_T \nabla^2 \delta({\bf r-r'}) \delta(t-t')
                                           \delta_{\alpha\beta} \ .
\end{eqnarray}
The equilibrium free energy functional has the form
\begin{equation} \label{GL}
{\cal F}  = \int d^3 r \ \left[ c(\nabla^2 \Phi)^2
          + g(\Phi) (\nabla \Phi)^2 + f(\Phi) +
          \frac{1}{2}{\bf j}_T^2 \right] \ .
\end{equation}
It has been shown [17-22]
that the {\it static} behavior of ternary amphiphilic systems is very
well described by this free energy functional, with ${\bf j}_T({\bf
r},t)$ integrated out, when the functions $f$ and $g$ are chosen to be
\begin{eqnarray}
g(\Phi) &=& b_0 + b_2 \Phi^2 \\
f(\Phi) &=& r_2 \Phi^2 + r_4 \Phi^4 + r_6 \Phi^6 \ ,
\end{eqnarray}
with $c>0$, $b_2>0$, and $r_6>0$. Three-phase coexistence between
oil-rich, water-rich and microemulsion phases is obtained for $r_4<0$,
$r_2>0$ and $b_0<0$.
As discussed above, we assume that the transport of the amphiphile is
much faster than all the other transport processes. Thus, the
deviation of the amphiphile concentration $\psi({\bf r}, t)$ from its
mean value $\bar\psi = <\psi({\bf r}, t)>$ does not appear explicitly
in our model. However, the average concentration $\bar\psi$ and the
properties of the amphiphile enter the model via the parameters of the
functions $f$ and $g$. The value of $g(\Phi)$ in the microemulsion,
$b_0$, for example, can be made negative by increasing the amphiphile
concentration or the strength of the amphiphile \cite{Lerczak}. Since
we are interested here only in the behavior of the microemulsion
phase, it is sufficient to retain only quadratic terms \cite{TeuStrey}
in the free energy functional (\ref{GL}), so that
\begin{eqnarray} \label{Gauss_g}
g(\Phi) &=& b_0 \\
\label{Gauss_f}
f(\Phi) &=& r_2 \Phi^2  \ .
\end{eqnarray}

\par The same model can also be used to describe sponge phases
\cite{RCOBNB,sponge,GoSch94}. In this case, the order parameter
$\Phi({\bf r}, t)$ is identified with the local concentration
difference between water on one side ("inside") and the other side
("outside") of the amphiphilic bilayer. Since we use a conserved order
parameter in (\ref{Langevin1}) and (\ref{Langevin2}), our results
apply to sponge phases in which the leakage time of water through the
bilayer is very large \cite{sponge_dyn}.

\section{Van-Hove Theory}

\par When non-linear contributions in the Langevin equations
(\ref{Langevin1}), (\ref{Langevin2}) are neglected, all structure
factors can be calculated easily \cite{Mazenko}. In this
approximation, the Fourier transform of Eq.(\ref{Langevin1}) reads,
for example,
\begin{equation} \label{Langevin_lin}
-i\omega \Phi({\bf k}, \omega) =
  - 2 \Gamma_\Phi k^2 \left( c k^4 + b_0 k^2 + r_2 \right)
                                            \Phi({\bf k}, \omega)
          + \zeta_\Phi({\bf k}, \omega) \ .
\end{equation}
Since (\ref{Langevin_lin}) is a linear relation between the random
variable $\Phi({\bf k}, \omega)$ and the noise $\zeta_\Phi({\bf k},
\omega)$, the noise correlation function (\ref{noise_phi}) immediately
implies that the dynamic water-water correlation function has the form
\begin{equation} \label{s_komega}
G_{\Phi\Phi}^{(0)}({\bf k},\omega) = \frac{2 \Gamma_\Phi k^2}
     {\omega^2 + \left[ \Gamma_\Phi k^2 \chi_0(k)^{-1} \right]^2 } \ ,
\end{equation}
where
\begin{equation}\label{statstruct}
\chi_0(k)^{-1} = 2 \left( c k^4 + b_0 k^2 + r_2 \right)
\end{equation}
is the static structure factor.
For $b_0^2 < 4cr_2$, the static correlation function in real space
is given by \cite{TeuStrey}
\begin{equation} \label{statcorr}
G_{\Phi\Phi}({\bf r}) = \frac{A}{r} e^{-r/\xi} \sin(qr)
\end{equation}
where $\xi$, with
\begin{equation} \label{xi}
\xi^{-2} = \frac{1}{2} \sqrt{\frac{r_2}{c}} + \frac{1}{4} \frac{b_0}{c}
\end{equation}
is the correlation length, and $2\pi/q$, with
\begin{equation} \label{domain}
q^{2} = \frac{1}{2} \sqrt{\frac{r_2}{c}} - \frac{1}{4} \frac{b_0}{c}
\end{equation}
is the characteristic domain size of coherent oil- and water-regions.
It follows from Eqs. (\ref{statstruct}) and (\ref{statcorr}) that
there are three important lines in the phase diagram
\cite{TeuStrey,GoSch90}. With decreasing $b_0$, the first is the
``disorder line", $b_0 = \sqrt{4 c r_2}$, where the correlation
function changes its behavior from a monotonic decay to damped
oscillations. The second line is the ``Lifshitz line", $b_0=0$, where
a peak in the structure factor begins to move to non-zero wavenumber.
Finally, the line $b_0 = - \sqrt{4 c r_2}$ is the spinodal to the
lamellar phase. Note that the shape of the static structure factor is
characterized by a single dimensionless parameter $q\xi$.

\par The intermediate scattering function, $G_{\Phi\Phi}({\bf k},t)$,
is obtained from Eq. (\ref{s_komega}) by Fourier transform, and reads
\cite{Mazenko}
\begin{equation} \label{intermediate}
G_{\Phi\Phi}^{(0)}({\bf k},t) = \chi_0(k)
\exp\left[-\Gamma_\Phi k^2 \chi_0(k)^{-1} t \right] \ .
\end{equation}
Both Eqs. (\ref{s_komega}) and (\ref{intermediate}) can be used to
define a characteristic frequency,
\begin{equation} \label{omegac}
\omega_c \equiv \Gamma_\Phi k^2 \chi_0(k)^{-1} \ .
\end{equation}
This characteristic frequency can be written in the scaling form
\begin{equation} \label{scale}
\omega_c = k^z \Omega_0( k\xi, k/q ) \ ,
\end{equation}
where the dynamic exponent $z=6$ and the scaling function
\begin{equation}
\Omega_0(x,y) = 2c\Gamma_\Phi \left[1 + 2(x^{-2} - y^{-2}) +
                 (x^{-2} + y^{-2})^2 \right]
\end{equation}
are obtained easily from Eq. (\ref{omegac}). The asymptotic behavior
of $\omega_c$ in various limits is listed in Table $I$. The inverse
scaling function, {\it i.e.} $k^z/\omega_c$, is shown in Fig. 1. Two
slow modes are present in a structured microemulsion, one for $k
\simeq 0$, the other for $k\simeq q$. In the first case, the
conservation of $\Phi$ causes the divergence of the relaxation time in
the limit $k\to 0$, as in any other binary fluid. In the latter case,
critical slowing down occurs as the transition to the lamellar phase
is approached; this slow mode is thus due to the increasing stability of
fluctuations with the wavevector $q$, which characterizes the structure
of the microemulsion.

\section{Perturbation Theory}

\par In order to treat the non-linear interactions, we employ a
field-theoretic approach \cite{BauJanWag,DeDomPel} using response
fields $\tilde\Phi$ and $\tilde {\bf j}_T$. The two-point correlation
and response functions $G_{\psi_i,\psi_j} (1,2) = <\psi_i(1)\psi_j(2)
>$ are given exactly by Dyson's equation,
\begin{equation}
\left[G_{\psi_i,\psi_j}(1,2) \right]^{-1} =
\left[G_{\psi_i,\psi_j}^{(0)}(1,2)\right]^{-1} -
\Sigma_{\psi_i,\psi_j}(1,2)
\end{equation}
in terms of the self-energies $\Sigma_{\psi_i,\psi_j}$, with $\psi_i
\in \{\Phi,\tilde\Phi, {\bf j}_T, \tilde {\bf j}_T \}$. Here,
$G_{\psi_i,\psi_j}^{(0)}(1,2)$ denotes the correlation and response
functions of the linearized theory. The inversion of the Dyson
equation yields in particular the correlation function
\begin{equation} \label{corr_phiphi}
G_{\Phi\Phi}({\bf k}, \omega) = \frac {2\Gamma_\Phi k^2 +
                                    \Sigma_{\tilde\Phi\tilde\Phi} }
  {| -i\omega + \Gamma_\Phi k^2 \chi_0(k)^{-1} -
                                    \Sigma_{\Phi\tilde\Phi} |^2 } \ .
\end{equation}
In the one-loop approximation, the self-energy
$\Sigma_{\Phi\tilde\Phi}({\bf k}, \omega)$ is given by \cite{SiHaHo}
\begin{equation} \label{sigma_phi2}
\Sigma_{\Phi\tilde\Phi}({\bf k}, \omega) = - g_0^2
\int \frac{d^3p}{(2\pi)^3} \ \frac{ \chi_0(p) }{ \chi_0(k) } \
\frac{ {\bf k}\cdot{\cal T}_{\bf p-k}\cdot{\bf k} }
{-i\omega + \Gamma_T({\bf p-k})^2
                    + \Gamma_\Phi p^2 \chi_0(p)^{-1} } \ ,
\end{equation}
with the projection operator
\begin{equation}
{\cal T}_{\bf k}^{\alpha\beta} = \delta_{\alpha\beta} -
                \frac{k_\alpha k_\beta} {k^2} \ .
\end{equation}
The Feynman diagrams are shown in Fig. 2a.
The second self-energy, $\Sigma_{\tilde\Phi\tilde\Phi}$, in Eq.
(\ref{corr_phiphi}) can be obtained from Eq. (\ref{sigma_phi2}) by
the fluctuation dissipation theorem, which leads to
\begin{equation} \label{FDT}
\Sigma_{\tilde\Phi\tilde\Phi}({\bf k}, \omega) =
     - 2 \chi_0(k) \  {\rm Re}\left\{
                \Sigma_{\Phi\tilde\Phi}({\bf k}, \omega) \right\} \ ,
\end{equation}
or directly from the loop expansion, just as
$\Sigma_{\Phi\tilde\Phi}$. The calculation of the other correlation
and response functions is described in the appendix.

\par To calculate the characteristic frequency $\omega_c$ in
perturbation theory, we define \cite{GraCates}
\begin{equation}
\omega_c = \frac{G_{\Phi\Phi}(k, t=0)}
          {\int_0^\infty dt G_{\Phi\Phi}(k, t)}
         = 2 \frac{ \int_{-\infty}^\infty \frac{d\omega}{2\pi}
              G_{\Phi\Phi}(k, \omega)}{G_{\Phi\Phi}(k, \omega=0) }
         = \frac{2\chi_0(k)}{G_{\Phi\Phi}(k, \omega=0)} \ .
\end{equation}
In the case of the van-Hove theory, this definition agrees with Eq.
(\ref{omegac}).

\par The results of van-Hove and perturbation theory for  $\omega_c$
are compared in Fig. 3. The dependence of $\omega_c$ on the wavevector
$k$ shown in Fig. 3a demonstrates that $\omega_c$ becomes larger (i.e.
the typical relaxation times becomes {\it smaller}) when the
non-linear interactions are taken into account. This shows that the
hydrodynamic flow speeds up the relaxation process of the
order-parameter correlations. This hydrodynamic relaxation mode must
be a flow of oil and water through their multiply connected
networks which are characteristic for a microemulsion.

\par Fig. 3b shows $\omega_c$ for $k=q$. This demonstrates again the
point we have made above for the van-Hove theory that there is a slow
mode, with a relaxation time which diverges as the spinodal of the
transition to the lamellar phase is approached, i.e. for $1/(q\xi)\to
0$. However, it can also be seen from Fig. 3b that the relaxation time
obtained from the perturbation theory remains finite in the limit
$q\xi\to\infty$. This is a disturbing result, since it can easily be
seen from Eqs. (\ref{GL}), (\ref{Gauss_g}) and (\ref{Gauss_f}) that
the static spinodal remains unchanged.

\par It is not difficult to find the origin of this problem. It is the
use of the van-Hove response function $G_{\tilde jj}^{(0)}$ of the
momentum density in Eq. (\ref{sigma_phi1}) for the self-energy,
compare Fig. 2a. In the van-Hove approximation, the response and
correlation functions of the momentum density contain no information
about the structure of the microemulsion, as can be seen from Eq.
(\ref{corr_jj}). To avoid this problem, all one-loop correlation and
reponse functions have to be evaluated self-consistently. Since this
requires a considerable numerical effort, we only use the one-loop
result for $G_{\tilde jj}$ and $G_{jj}$ in Eq. (\ref{sigma_phi1}),
which is the dominant contribution. With this modification, the
characteristic freqency for $k=q$ now correctly vanishes at the
spinodal, as can be seen in Fig. 3c.

\par The dynamic correlation function $G_{\Phi\Phi}({\bf k}, \omega)$
as a function of $\omega$ for three different values of the wavevector
$k$ is shown in Fig. 4. Note that for $k \simeq q/2$ a peak develops
for $\omega>0$ in the limit of large $q\xi$, i.e. for strongly
structured microemulsions. This peak indicates that the correlation
function {\it oscillates} in {\it time}. Indeed, the Fourier transform
of the data shown in Fig. 4b has temporal oscillations, see Fig. 5b. For
$k > q$, the peak is much broader, so that there are still
oscillations, but with a rapid decay, see Fig. 5a. Finally, for $k \ll
q$, the peak at finite $\omega$ is absent, and the correlations
function for large times decays monotonically, as can be seen in Fig.
5c. The peak at finite $\omega$ and the temporal oscillations of the
{\it dynamic} correlation function are reminiscent of the peak
at finite $k$ and the spatial oscillations of the {\it static}
correlation function. However, while the latter has been interpreted
convincingly in terms of the microemulsion structure
\cite{TeuStrey,VoBiKa}, the mechanism of the former is unclear.

\par Finally, we show the dynamic correlation function
$G_{\Phi\Phi}({\bf k}, \omega)$ as a function of the wavevector $k$
for fixed dimensionless freqency $\omega/\Gamma_{\phi}=0.3$ in Fig. 6.
This function vanishes in the limit $k\to 0$ due to the conservation
of the order parameter.

\section{Channel Flow}

\par To understand the temporal oscillations of the correlation
function in microemulsions, we follow an idea which was used to
explain the dynamics of spinodal phase separation in the bicontinuous
regime \cite{Siggia}. It is argued in that case that the relaxation is
dominated by the flow through the channels; this leads to a growth law
for the coarsening of the bicontinuous structure, which is linear in
time $t$, rather than the $t^{1/3}$ behavior observed for droplets
\cite{LifSlo}. The force which drives the flow through the channels is
the interfacial tension.

\par To derive an equation which describes the time dependence
observed in microemulsions, we start from the linearized Navier-Stokes
equation \cite{Landau} for the velocity field ${\bf v}({\bf r},t)$,
\begin{equation} \label{NSE}
\frac{\partial}{\partial t} {\bf v} = - \frac{1}{\rho} \nabla p +
\frac{\eta}{\rho} \nabla^2 {\bf v}
\end{equation}
where $\rho$ is the mass density, and $\eta$ is the viscosity.
Consider a spherical droplet of radius $L$ within the microemulsion
phase, in which the concentrations of oil and water deviate slightly
from their average values, as shown schematically in Fig. 7. This
implies that the cross-section of the oil- and water channels in this
region is somewhat larger (or smaller) than on average. The easiest
way for the system to get rid of the extra oil and/or water is to let
it flow through the channels to the outside of the droplet. We now
assume that the flow through this multiply connected network of a
microemulsion is similar to the flow through a cylindrical tube of
length $L$. For a cylinder in the $z$-direction, the flow field is
approximated by Poiseuille flow,
\begin{equation} \label{Poiseuille}
{\bf v}({\bf r},t) = \alpha(t)[R(t)^2 - r^2] {\bf e}_z
\end{equation}
where $R(t)$ is the radius of the tube and $\alpha(t)$ describes the
magnitude of the velocity. Further, we employ the Laplace equation to
relate the pressure gradient to the tube geometry and to the
interfacial tension $\sigma$,
\begin{equation} \label{laplace}
\nabla p \simeq -\frac{\Delta p}{L} {\bf e}_z =
                          - \frac{\sigma}{R L}{\bf e}_z
\end{equation}
for a tube of length $L$. The Navier-Stokes equation (\ref{NSE})
then implies that the average velocity
(averaged over the cross-section of the cylinder),
\begin{equation} \label{ave_velo}
\bar v(t) = \frac{1}{R(t)} \int_0^{R(t)} v(r,t) dr
\end{equation}
is determined by
\begin{equation}
\frac{\partial}{\partial t} \bar v + {\bar v}\frac{\dot R}{R} =
   \frac{\sigma}{\rho R L}
       - \frac{6\eta}{\rho} \frac{\bar v}{R^2} \ ,
\end{equation}
where $\dot R = dR/dt$.
Note that it follows immediately from Eqs. (\ref{Poiseuille}) and
(\ref{ave_velo}) that $\bar v = \frac{2}{3} \alpha R^2$.
Since for an incompressible fluid
\begin{equation}
\bar v(t) = - \frac{2L}{R(t)} \dot R(t)
\end{equation}
we finally arrive at
\begin{equation} \label{rdot}
\ddot R = - \frac{\sigma}{2L^2\rho} -
\frac{6\eta}{\rho R^2}  \dot R + \frac{2}{R} {\dot R}^2  \ .
\end{equation}
The analysis presented so far is not specific to bicontinuous
microemulsions, but equally well applies to the case of bicontinuous
phase separation. The specificity of amphiphilic systems comes in when
we consider the interfacial tension $\sigma$. It is well known that
the interfacial tension between the oil-rich and the water-rich phase
in these systems is very small. What we really need in Eqs.
(\ref{laplace}) and (\ref{rdot}), however, is not the tension of the
oil/water interface, but the driving force for a change in the tube
radius. It has been shown in Refs. \cite{GolLub} and \cite{GoKrausI}
that in the {\it lamellar phase} the tension vanishes identically for
the equilibrium separation of the interfaces. However, if the spacing
between the lamellae is increased (decreased), a negative (positive)
tension develops \cite{GoKrausI}, which drives the interfaces back to
their equilibrium distance. We assume here that the same is true
in the microemulsion. For small deviations from the equilibrium radius
$R_0$, we can thus write
\begin{equation} \label{tension}
\sigma(R) = \tilde p (R-R_0) + O((R-R_0)^2)
\end{equation}
with a positive constant $\tilde p$, and
\begin{equation}
R(t) = R_0 + \epsilon(t) \ .
\end{equation}
By inserting these results into Eq. (\ref{rdot}), we find that
$\epsilon(t)$ is determined by
\begin{equation} \label{tube_osc}
\ddot \epsilon +
    6\frac{\eta}{\rho R_0^2} \dot \epsilon
        + \frac{\tilde p}{2L^2\rho} \epsilon = 0
\end{equation}
to leading order in $\epsilon$. Thus, $\epsilon(t)$ satisfies the
differential equation of a damped harmonic oscillator. The radius
oscillates if the damping is small enough, i.e. if
\begin{equation} \label{cond_osc1}
\frac{\tilde p}{2L^2\rho} >
         \left(3\frac{\eta}{\rho R_0^2} \right)^2  \ .
\end{equation}
Thus, there should be oscillations if the viscosity $\eta$ of oil or
water are small enough. We want to emphasize that the $R$-dependence
of the tension $\sigma$ in Eq. (\ref{tension}) is unrelated to a
stretching of the amphiphilic monolayer (i.e. a change in the area per
headgroup), since we are assuming in our model (\ref{GL}) that the
amphiphile concentration samples its full equilibrium distribution on
a time scale rapid compared to changes in the oil- and
water-concentrations.

\par With these results, we can now predict the behavior of the
dynamic water-water correlation function. We identify the tube radius
$R_0$ with $\pi/q$, and the tube length $L$ with $\pi/k$. The
viscosity in our Ginzburg-Landau model is given by $\eta =
\rho\Gamma_T$ \cite{KadMartin}. In terms of the variables $q$, $k$ and
$\Gamma_T$, the condition (\ref{cond_osc1}) then reads
\begin{equation} \label{cond_osc2}
k^2 > a_0 \Gamma_T^2 q^4 \ ,
\end{equation}
where $a_0 = 18 \rho/(\pi^2 \tilde p)$. Here, both the parameter
$\tilde p$ and the mass density $\rho$ depend on the coefficients $c$,
$b_0$ and $r_2$ in the free energy functional (\ref{GL}), but {\it
not} on any dynamic coefficients in Eqs. (\ref{Langevin1}),
(\ref{Langevin2}). While the variation of $\rho$ is probably small,
$\tilde p$ can vary considerably. Its functional dependence can be
obtained from a calculation of the interfacial tension
\cite{GoKrausI}. In the following discussion, we will consider a fixed
set of parameters of the free energy functional (\ref{GL}),
(\ref{Gauss_g}), (\ref{Gauss_f}), so that $a_0$ and $q$ are constant.

\par In case the inequality (\ref{cond_osc2}) is satisfied, the
solution of Eq. (\ref{tube_osc}) then implies that the asymptotic
decay of the correlation function should have the form
\begin{equation} \label{asympt}
G_{\Phi\Phi}({\bf k}, t) = B \exp(-t/\tau) \cos(\nu(k) t + \varphi)
\end{equation}
where
\begin{eqnarray} \label{relax}
\tau &=& \frac{a_1}{\Gamma_T q^2} \\
\label{frequency}
\nu(k)^2 &=& \frac{a_2}{a_0} k^2 - a_3 \Gamma_T^2 q^4
\end{eqnarray}
with constants $B$, $\varphi$, and $a_1$, $a_2$, $a_3$. We want to
emphasize that the relaxation of regions with increased oil- or
water-concentrations is facilitated by the flow mechanism on length
scales larger than the diameter of a tube, i.e. for $k\ll q$. This is
the regime where we expect our result (\ref{asympt}) with
(\ref{cond_osc2}) to describe the dynamics correctly. On the other
hand, on length scales smaller than the tube radius, i.e. for $k>q$,
the dynamics should be determinded by undulations of the microscopic
oil/water interfaces, or by the fluctuations within the channels,
which are ignored in our derivation.

\par Our predictions (\ref{cond_osc2}) and (\ref{asympt}) with
(\ref{relax}) and (\ref{frequency}) for the correlation function can
now be compared with the result obtained from the time-dependent
Ginzburg-Landau theory. First, note that the inequality
(\ref{cond_osc2}) implies that there should be no oscillations for
sufficiently small $k$, in agreement with the result shown in Fig. 5c.
To extract the relaxation time and the oscillation frequency from the
perturbation theory, we fit the inverse correlation function, for
small $\omega$ (up to the peak position) to a fourth-order polynomial,
\begin{equation}
G_{\Phi\Phi}({\bf k}, \omega)^{-1} = \mu_0(k) + \mu_2(k) \omega^2
                + \mu_4(k) \omega^4 + O(\omega^6)
\end{equation}
and then use the equivalent of Eqs. (\ref{xi}) and (\ref{domain}) to
calculate $\tau$ and $\nu$. We have compared the results in a few cases
with the explicit Fourier transform, $G_{\Phi\Phi}({\bf k}, t)$, and
have found good agreement. The relaxation time $\tau$ and the
oscillation frequency $\nu$ as functions of $k$ and $\Gamma_T$ are
shown in Fig. 8 and 9. It can be seen that in the regime $k \ll q$
there is semi-quantitative agreement with predictions (\ref{relax})
and (\ref{frequency}).  The linear dependence of $1/\tau$ on
$\Gamma_T$ and of $\nu^2$ on  $\Gamma_T^2$ is found to be satisfied
very nicely, see Figs. 8a and 9a. However, there is a considerable
$k$-dependence of $\tau$, see Fig. 8b, although from (\ref{relax}) it
is expected to be constant. In the regime $k {> \atop \sim} q$, where
we do not expect our arguments leading to Eqs. (\ref{relax}) and
(\ref{frequency}) to apply, the behavior of $\tau$ and $\nu$ as a
function $k$ is indeed qualitatively different. The relaxation time is
found to have a maximum at $k=q$, while the oscillation frequency
almost vanishes, see Figs. 8b and 9b. Finally, for $k>q$, the
relaxation time decreases rapidly.

\section{Sponge Phases}

\par It has been pointed out in Refs. \cite{RCOBNB,sponge} that in
sponge phases the order parameter $\Phi$ and its fluctuations cannot
be observed directly, since water on one side of the membrane cannot
be distinguished from water on the other side. The order parameter
fluctuations can been seen indirectly, however, via the fluctuations
of the amphiphile density $\psi({\bf r}, t)$.

\par Although our model (\ref{GL}) does not contain any explicit
amphiphile degrees of freedom, we can nevertheless calculate an
amphiphile correlation function, if we assume that most of the
amphiphile is located at the $\Phi({\bf r},t) = 0$ surfaces. Here, we
assume again that the amphiphile concentration equilibrates
instantaneously with relatively slow changes of the order-parameter
field $\Phi({\bf r},t)$. In this case, the amphiphile correlation
function is given by
\begin{equation}
G_{film}({\bf r}, t) =
N_0 < \delta(\Phi({\bf r},t)) \ \delta(\Phi({\bf 0},0)) > \ ,
\end{equation}
with a normalization factor $N_0$, which is chosen such that
$\lim_{r\to\infty} \lim_{t\to\infty} G_{film}({\bf r}, t) = 1$.
The average can be calculated easily for Gaussian fluctuations, where
one finds \cite{PierMarc,GoGoos94}
\begin{equation} \label{film_corr_G}
G_{film}({\bf r}, t) = \frac{N_0}{2\pi}
     \left[ <\Phi({\bf 0},0)^2>^2
            - <\Phi({\bf r},t)\Phi({\bf 0},0))>^2 \right]^{-1/2} \ .
\end{equation}
For not too small $r$ or $t$, this result can be expanded to give
\begin{equation} \label{film_asym}
G_{film}({\bf r}, t) - G_{film}(r=\infty, t=\infty)
            \sim \ <\Phi({\bf r},t)\Phi({\bf 0},0))>^2 \ .
\end{equation}
Interestingly, in the static case Eq. (\ref{film_asym}) is just the
result obtained from the two-order-parameter model, where the
amphiphile density is included explicitly \cite{RCOBNB}. We thus
identify the expression (\ref{film_asym}) with the amphiphile
correlation function, $G_{\psi\psi}({\bf r}, t)$.

\par The results for the amphiphile correlation function, calculated
with the van-Hove and the one-loop approximation for the
order-parameter correlations, Eqs. (\ref{corr_phiphi}) and
(\ref{sigma_phi2}), are shown in Fig. 10 as a function of the frequency
$\omega$ for $k=q/2$. It can be seen that the scattering intensity in
film contrast does not have a peak at finite $\omega$, but develops a
shoulder at the same value of $\omega$, where the order-parameter
correlation function has its peak.

\par We want to point out that a very similar behavior is found when
the expression (\ref{film_asym}) is used to calculate the {\it static}
film correlation function \cite{GSreview}. Only when additional
coupling terms in a free energy functional for the {\it two} order
parameters $\Phi$ and $\psi$ are taken into account, a peak at finite
wavevector $k$ appears \cite{GoSch94}. Thus, we expect that a
time-dependent Ginzburg-Landau model for $\Phi$ and $\psi$ will
show a peak of $G_{\psi\psi}({\bf k},\omega)$ at a finite frequency
$\omega$.

\par By a numerical Fourier transformation of the data of Fig. 10, we
obtain the behavior of the intermediate scattering function in film
contrast, $G_{\psi\psi}({\bf k}, t)$, as a function of time $t$. The
result is shown in Fig. 11 for the same wavevector $k=q/2$ for which
the order-parameter correlation function oscillates in time, compare
Fig. 5b. We find again an {\it oscillatory} behavior, with a
characteristic frequency which is roughly the same as the
characteristic frequency $\nu$ of $G_{\Phi\Phi}({\bf k}, t)$, but with
a considerably smaller relaxation time.

\par Note that formally there is a strong similarity with the spatial
behavior of the {\it static} correlation, where one finds with Eqs.
(\ref{statcorr}) and (\ref{film_asym})
\begin{equation}
G_{\psi\psi}({\bf r}) \sim \frac{A^2}{r^2} e^{-2r/\xi} \sin(qr)^2 \ ,
\end{equation}
a function which shows the same characteristics as a function of $r$
as $G_{\psi\psi}({\bf k}, t)$ does as a function of $t$. In
particular, $G_{\psi\psi}({\bf k}, t) \ge G_{\psi\psi}({\bf k},
t=\infty)$. We want to emphasize that these results of the
time-dependent Ginzburg-Landau theory agree with the arguments
presented in Section 5. From these arguments, it follows immediately
that $G_{\psi\psi}({\bf k}, t)$ should oscillate in the same intervall
of wavevectors as the order-parameter correlation function itself.
Also, the oscillation period of the two correlation functions should
be the same.

\section{Summary and Conclusions}

\par We have studied the dynamic behavior of microemulsion and sponge
phases in thermal equilibrium. For a time-dependent Ginzburg-Landau
model, we have analysed the dynamic structure factor $G({\bf k},
\omega)$, both in bulk and in film contrast, which can be measured,
for example, by inelastic neutron scattering experiments in
microemulsions. Our calculation shows that for systems with a
sufficiently small viscosity $\eta$, or with a sufficiently large
domain size $2\pi/q$ of coherent oil- and water-domains,
$G_{\Phi\Phi}({\bf k}, \omega)$ for fixed wavenumbers $k$ of order
$q/2$ develops a peak or shoulder at finite $\omega$. This peak implies
that the intermediate scattering function, $G_{\Phi\Phi}({\bf k}, t)$,
oscillates in time. To understand this surprising result, we have
studied the flow through tubes in a very simple, cylindrical geometry,
employing the linearized Navier-Stokes equation. Under the assumption
that the interfacial tension is radius-dependent, and vanishes
identically for the equilibrium radius of the tube, we were indeed
able to reproduce the oscillatory behavior of $G_{\Phi\Phi}({\bf k},
t)$. Furthermore, we found that the dependence of both the relaxation
time and the oscillation frequency on the viscosity in the two
approaches agrees very well, while the dependence on the wavenumber
$k$ is qualitatively correct. Thus, we believe that the existence of
temporal oscillations in microemulsions is well established.

\par The behavior of the amphiphile correlation function,
$G_{\psi\psi}({\bf k}, \omega)$, is found to be very similar. In the
range of wavevectors, where $G_{\Phi\Phi}({\bf k}, \omega)$ has a peak
at finite frequency $\omega_0$, $G_{\psi\psi}({\bf k}, \omega)$
develops a shoulder, which appears also at frequency $\omega_0$. The
intermediate scattering intensity is found to oscillate in time. The
characteristic time scale of these oscillations is the same in film
and in bulk contrast.

\par We want to emphasize that the two approaches we have used should
apply in two different limits. The Ginzburg-Landau approach is based
on the assumption that the order parameter fields vary slowly in space
and time, i.e. it applies to the weak segregation limit. The
hydrodynamic flow through a tube with hard walls, on the other hand,
should describe the behavior of the strong segregation limit. Since
the same behavior is found in both limits, it should also be found in
intermediate cases.

\par The mechanism, which leads to the oscillations of the
intermediate scattering function in bulk and film contrast, can be
summarized as follows. In a microemulsion or sponge phase, there is
some prefered average distance between the amphiphilic membranes. If
by thermal fluctuations this distance increases or decreases in some
region of these phases, there is a force, the interfacial tension,
which drives the system back to the thermal average. It is negative
(positive) for large (small) distances between membranes, and thus
tends to increase (decrease) the interfacial area. A restoring force
itself is not sufficient to produce oscillations. We also need a
moment of inertia, which drives the system through the point where the
force vanishes; in our model, this term is provided by hydrodynamic
momentum conservation. And we need a sufficiently small friction
coefficient; in the microemulsion or sponge phase, this can be
achieved by a sufficiently small viscosity, or by a sufficiently large
diameter of the tubes, through which the flow moves back and forth.
Thus, oscillations can only be observed for strongly structured and
swollen microemulsion or sponge phases.

\par We want to conclude with a short dicussion of the dynamic
behavior of the amphiphile. We have already pointed out that the
radius-dependence of the interfacial tension is not due to a
stretching of the amphiphile film, since in our model the amphiphile
always samples its equilibrium configurations. Let us now consider the
opposite limit where the dynamics of the amphiphile is very slow.
When the oil- or water-concentration in a drop of radius $L\gg \pi/q$
{\it decreases}, the interfacial area must increase. Since the amphiphile
dynamics is slow, its concentration in the drop can be assumed to be
approximately constant. This implies that  the amphiphile density {\it
within} the monolayer decreases, leading to an increase of the area
per headgroup and thus to a  {\it positive} interfacial tension.
However, this is the situation where we already have a positive
tension in our model (\ref{Langevin1}), (\ref{Langevin2}), (\ref{GL})
with an equilibrated amphiphile. Thus, if the amphiphile dynamics is
slow, the stretching of the monolayer {\it enhances} the radius
dependence of the interfacial tension, and thereby the oscillations.

\bigskip
\noindent
{\bf Acknowledgements:}
Helpful discussions with R. Hausmann, D.M. Kroll and H. Wagner are
gratefully acknowledged. This work was supported in part by the
Deutsche Forschungsgemeinschaft through Sonderforschungsbereich 266.

\newpage

\section{Appendix: Correlation and Response Functions}

\par The inversion of the Dyson equation leads to the
correlations functions $G_{\Phi\Phi}({\bf k}, \omega)$
given in Eq. (\ref{corr_phiphi}) and
\begin{equation} \label{corr_jj}
G_{j j}({\bf k}, \omega) =  \frac{2\Gamma_T k^2 +
                                    \Sigma_{\tilde j \tilde j} }
  {| -i\omega + \Gamma_T k^2 - \Sigma_{j \tilde j} |^2 } \ ,
\end{equation}
as well as to the response functions
\begin{equation} \label{resp_phiphi}
G_{\tilde \Phi \Phi}({\bf k}, \omega) =
      \frac{1}{-i\omega + \Gamma_\Phi k^2 \chi_0(k)^{-1} -
                         \Sigma_{\Phi\tilde\Phi} }
\end{equation}
and
\begin{equation} \label{resp_jj}
G_{\tilde j j}({\bf k}, \omega) =
\frac{1}{-i\omega + \Gamma_T k^2 - \Sigma_{j \tilde j} } \ .
\end{equation}
The correlation and resonse functions, $G_{\psi_i, \psi_j}^{(0)}
({\bf k}, \omega)$ of the linearized theory can be obtained easily
from these expressions by ignoring the self-energy contributions.

\par In the one-loop approximation, the self-energy
$\Sigma_{\Phi\tilde\Phi}({\bf k}, \omega)$ is given by
\begin{eqnarray} \label{sigma_phi1}
\Sigma_{\Phi\tilde\Phi}({\bf k}, \omega) &=&
g_0^2 \int \frac{d^3p}{(2\pi)^3} \int \frac{d\omega'}{2\pi} \
{\bf k}\cdot{\cal T}_{\bf p-k}\cdot{\bf k}  \\
& & {\hskip 1.5 true cm}
 \biggl\{ \left[ \chi_0(p)^{-1} - \chi_0(k)^{-1} \right]
     G_{\tilde jj}^{(0)}({\bf p-k}, \omega-\omega')
     G_{\Phi\Phi}^{(0)}({\bf p}, \omega')  \nonumber \\
 & & {\hskip 3.5 true cm}
 - G_{jj}^{(0)}({\bf p-k}, \omega-\omega')
      G_{\tilde \Phi\Phi}^{(0)}({\bf p}, \omega') \biggr\} \ ,\nonumber
\end{eqnarray}
see Fig. 2a. The explicit form has already been presented in Eq.
(\ref{sigma_phi2}).

\par The one-loop expression for $\Sigma_{j\tilde j}$ is given by
\begin{eqnarray} \label{sigma_j1}
\Sigma_{j\tilde j}({\bf k}, \omega) &=&
- g_0^2 \int \frac{d^3p}{(2\pi)^3} \int \frac{d\omega'}{2\pi} \
{\bf p}\cdot{\cal T}_{\bf k}\cdot{\bf p}  \\
& & {\hskip 1.5 true cm}
  \left[ \chi_0(p)^{-1} - \chi_0({\bf p-k})^{-1} \right]
     G_{\tilde \Phi\Phi}^{(0)}({\bf p}, \omega')
     G_{\Phi\Phi}^{(0)}({\bf p-k}, \omega-\omega') \ , \nonumber
\end{eqnarray}
see Fig. 2b, or equivalently \cite{SiHaHo}
\begin{eqnarray} \label{sigma_j2}
\Sigma_{j\tilde j}({\bf k}, \omega) &=&
- g_0^2 \int \frac{d^3p}{(2\pi)^3} \
 \left[ \chi_0(p)^{-1} - \chi_0({\bf p-k})^{-1} \right]
                                  \chi_0({\bf p-k})  \\
& & {\hskip 1.5 true cm}
    \frac{{\bf p}\cdot{\cal T}_{\bf k}\cdot{\bf p} }
      {-i\omega + \Gamma_\Phi p^2 \chi_0(p)^{-1} +
      \Gamma_\Phi({\bf p-k})^2 \chi_0({\bf p-k})^{-1} } \ . \nonumber
\end{eqnarray}
Finally, the self-energy, $\Sigma_{\tilde j\tilde j}$, in Eq.
(\ref{corr_jj}) can again be obtained from Eq.
(\ref{sigma_j2}) by the fluctuation dissipation theorem, which leads
to
\begin{equation}
\Sigma_{\tilde j\tilde j}({\bf k}, \omega) =
     - 2  {\rm Re}\left\{
                \Sigma_{j \tilde j}({\bf k}, \omega) \right\} \ ,
\end{equation}
or directly from the loop expansion, just as $\Sigma_{j \tilde j}$.

\newpage

\begin{tabular}{|c|l|}
\hline
 regime & \ \ \ \ \ \ \ \ \ $\omega_c$ \\
\hline
\hline
$\ k \gg \xi^{-1},q \ $ & $\ 2c\Gamma_\Phi k^6$ \\
\hline
$\ k \ll \xi^{-1},q\ $ & $\
                         2c\Gamma_\Phi (\xi^{-2} + q^2)^2 k^2$\ \ \\
\hline
$\ q \ll k \ll \xi^{-1}\ $ &\ $ 2c\Gamma_\Phi \xi^{-4} k^2$ \\
\hline
$\ \xi^{-1} \ll k \ll q\ $ &\ $ 2c\Gamma_\Phi q^4 k^2$ \\
\hline
$\ \xi^{-1}=0,\ k\simeq q\ $ &\ $  8c\Gamma_\Phi q^4 (k-q)^2$ \\
\hline
\end{tabular}

\bigskip
\bigskip
\noindent
Table $I$: Asymptotic behavior of the characteristic frequency
$\omega_c$ in the van-Hove approximation.

\newpage

\vskip .5cm
{\Large{\bf Figure Captions}}

\vskip 0.5cm

Figure 1: The inverse scaling function $\Omega_0(k\xi,k/q)^{-1} =
k^z/\omega_c$ in the van-Hove approximation.

\vskip 0.5cm

Figure 2: Feynman diagrams which contribute to one-loop order to  the
self-energies (a) $\Sigma_{\Phi\tilde \Phi}$ and (b) $\Sigma_{j\tilde
j}$. Here, the straight lines represent the physical fields, the wavy
lines the response fields. The solid lines denote order parameter
propagators, the lines with bars momentum propagators.
The diagrams contain the vertex contributions  ${\bf u}={\bf
p}\chi_0({\bf p})^{-1} - {\bf k} \chi_0({\bf k})^{-1}$ and
${\bf v}={\bf p} \left[\chi_0({\bf p-k})^{-1} - \chi_0({\bf p})^{-1}
\right]$.

\vskip 0.5cm

Figure 3: The results of van-Hove (dashed lines) and perturbation
theory (full lines) for $\omega_c$, (a) as a function of $k/q$ for fixed
$q\xi=8.888$, (b)(c) as a function of $q\xi$ for fixed $k=q$. (c)  The
behavior of $\omega_c$ for large $q\xi$ in detail. Here, the
dashed-dotted line is calculated with the one-loop approximation for
$G_{jj}$ and $G_{\tilde jj}$. The parameters in Eqs.
(\ref{Langevin1}), (\ref{Langevin2}),  (\ref{GL}), (\ref{Gauss_g}) and
(\ref{Gauss_f}) are $\Gamma_\Phi=1$, $\Gamma_T=1$, $g_0=2$, $c=1$, and
$r_2=1$.

\vskip 0.5cm

Figure 4: The dynamic correlation function $G_{\Phi\Phi}({\bf k},
\omega)$ as a function of the scaled frequency $\omega/\Gamma_\Phi$
for three values of the wavevector, (a) $k=1.3 q$, (b) $k=q/2$, and
(c) $k=q/10$. The dashed line is the van-Hove result, the full line
the result of the perturbation theory. The parameters in Eqs.
(\ref{Langevin1}), (\ref{Langevin2}), (\ref{GL}), (\ref{Gauss_g}) and
(\ref{Gauss_f}) are $\Gamma_\Phi=1$, $\Gamma_T=0.3$, $g_0=5$, $c=1$,
$r_2=1$, and $q\xi=8.888$.

\vskip 0.5cm

Figure 5: The intermediate scattering function $G_{\Phi\Phi}({\bf k},
t)$ as a function of the scaled time $t\Gamma_\Phi$ for three values
of the wavevector, (a) $k=1.3 q$, (b) $k=q/2$, and (c) $k=q/10$. The
parameters are the same  as in Fig. 4.

\vskip 0.5cm

Figure 6: The dynamic correlation function $G_{\Phi\Phi}({\bf k},
\omega)$ as a function of the scaled wavevector $k/q$ for
$\omega/\Gamma_{\Phi} = 0.3$. The dashed line is the van-Hove result,
the full line the result of the perturbation theory. The parameters
are the same as in  Fig. 4.

\vskip 0.5cm

Figure 7: A spherical droplet of radius $L$ (marked by the dashed
line) within the microemulsion phase, in which the concentrations of
oil (dark grey regions) and water (white regions) deviate from their
average values (schematically).

\vskip 0.5cm

Figure 8: The relaxation time $\tau$ of the asymptotic form
(\ref{asympt}) of the intermediate scattering function
$G_{\Phi\Phi}({\bf k}, t)$, as obtained from pertubation theory. (a)
$1/\tau$ as a function of $\Gamma_T$ for $k=q/2$ (full line) and
$k=q/3$ (dashed line). (b) $\tau$ as a function of $k$ for
$\Gamma_T=0.2$.  The parameters in Eqs. (\ref{Langevin1}),
(\ref{Langevin2}), (\ref{GL}), (\ref{Gauss_g}) and (\ref{Gauss_f}) are
$\Gamma_\Phi=1$, $\Gamma_T=0.3$, $g_0=5$, $c=1$, $r_2=1$, and
$q\xi=19.97$.

\vskip 0.5cm

Figure 9: The oscillation frequency $\nu$ of the asymptotic form
(\ref{asympt}) of the intermediate scattering function
$G_{\Phi\Phi}({\bf k}, t)$, as obtained from pertubation theory. (a)
$\nu^2$ as a function of $\Gamma_T^2$ for $k=q/2$ (full line) and for
$k=q/3$ (dashed line). (b) $\nu^2$ as a function of $k^2$ for
$\Gamma_T=0.2$. The other parameters are the same as in Fig. 8.

\vskip 0.5cm

Figure 10: The dynamic amphiphile correlation function
$G_{\psi\psi}({\bf k}, \omega)$ as a function of the scaled frequency
$\omega/\Gamma_\Phi$, for wavevector $k=q/2$. The dashed line is the
van-Hove result, the full line the result of the perturbation theory.
The parameters are the same as in Fig. 4.

\vskip 0.5cm

Figure 11: The intermediate scattering function in film contrast,
$G_{\psi\psi}({\bf k}, t)$, as a function of the scaled time
$t\Gamma_\Phi$, for wavevector $k=q/2$. (a) van-Hove result (dashed
line) and the result of the perturbation theory (full line). (b) A
magnification of $G_{\psi\psi}({\bf k}, t)$ (perturbation theory) on
intermediate time scales. The parameters are the same as in Fig. 4.


\newpage

\begin{thebibliography}{99}

\bibitem{books} J. Meunier, D. Langevin, and N. Boccara (eds.), {\it
Physics of Amphiphilic Layers}, Springer Proceedings in Physics {\bf
21} (Springer, Berlin, 1987); R. Lipowsky, D. Richter, and K. Kremer
(eds.), {\it The Structure and Conformation of Amphiphilic Membranes}
(Springer, Berlin, 1992); W.M. Gelbart, D. Roux, and A. Ben-Shaul
(eds.), {\it Micelles, Membranes, Microemulsions, and Monolayers}
(Springer, Berlin, to be published).

\bibitem{GSreview} G. Gompper and M. Schick, in {\it Phase Transitions
and Critical Phenomena}, Vol. 16, edited by C. Domb and J. Lebowitz
(Academic Press, London, 1994).

\bibitem{Porte} G. Porte, J. Phys. Cond. Matt. {\bf 4}, 8649 (1992),
and references therein.

\bibitem{sponge} D. Roux, C. Coulon, and M.E. Cates, J. Phys. Chem.
{\bf 96}, 4174 (1992).

\bibitem{Chowdhury} D. Chowdhury, J. Phys. Cond. Matt. {\bf 6}, 2435
(1994), and references therein.

\bibitem{SchBloss} F. Schmid and R. Blossey, preprint.

\bibitem{GraBaCates} R. Granek, R.C. Ball, and M.E. Cates, J. Phys. II
France {\bf 3}, 829 (1993).

\bibitem{Shell} S. Karaborni, N.M. van Os, K. Esselink, and P.A.J.
Hilbers, Langmuir {\bf 9}, 1175 (1993).

\bibitem{Stauffer} D. Stauffer, N. Jan, and R.B. Pandey, Physica A{\bf
198}, 401 (1993); N. Jan and D. Stauffer, J. Phys. I France {\bf 4},
345 (1994).

\bibitem{HoHa} P.C. Hohenberg and B.I. Halperin, Rev. Mod. Phys. {\bf
49}, 435 (1977).

\bibitem{Mazenko} G.F. Mazenko, in {\it Correlation Functions and
Quasiparticle Interactions in Condensed Matter}, edited by J.W.
Halley (Plenum Press, New York, 1978).

\bibitem{sponge_dyn} R. Granek, M.E. Cates, and S. Ramaswamy, Europhys.
Lett. {\bf 19}, 499 (1992).

\bibitem{Kawasaki} K. Kawasaki, in {\it Phase Transitions and Critical
Phenomena}, Vol. 5a, edited by C. Domb and M.S. Green (Academic Press,
London, 1976).

\bibitem{KrollRuh} D.M. Kroll and J.M. Ruhland, Phys. Lett. {\bf 80A},
45 (1980).

\bibitem{GomHen} G. Gompper and M. Hennes, Europhys. Lett. {\bf 25},
193 (1994).

\bibitem{SiHaHo} E.D. Siggia, B.I. Halperin, and P.C. Hohenberg, Phys.
Rev. B{\bf 13}, 2110 (1976).

\bibitem{TeuStrey} M. Teubner and R. Strey, J. Chem. Phys. {\bf 87},
3195 (1987).

\bibitem{GoSch90} G. Gompper and M. Schick, Phys. Rev. Lett. {\bf
65}, 1116 (1990).

\bibitem{GoHoSch} G. Gompper, R. Ho\l yst, M. Schick, Phys. Rev. A
{\bf 43}, 3157 (1991).

\bibitem{GoZsch} G. Gompper and S. Zschocke, Europhys. Lett. {\bf 16},
731 (1991); Phys. Rev. A {\bf 46}, 4836 (1992).

\bibitem{GoKrausI} G. Gompper and M. Kraus, Phys. Rev. E{\bf 47},
4289 (1993).

\bibitem{GoKrausII} G. Gompper and M. Kraus, Phys. Rev. E {\bf 47},
4301 (1993).

\bibitem{Lerczak} J. Lerczak, M. Schick, and G. Gompper, Phys. Rev.
A{\bf 46}, 985 (1992).

\bibitem{RCOBNB} D. Roux, M.E. Cates, U. Olsson, R.C. Ball, F. Nallet,
and A.M. Bellocq, Europhys. Lett. {\bf 11}, 229 (1990).

\bibitem{GoSch94} G. Gompper and M. Schick, Phys. Rev. E{\bf 49}, 1478
(1994).

\bibitem{BauJanWag} R. Bausch, H.K. Jansen, and H. Wagner, Z. Phys.
B{\bf 24}, 113 (1976).

\bibitem{DeDomPel} C. de Dominicis and L. Peliti, Phys. Rev. B{\bf
18}, 353 (1978).

\bibitem{GraCates} R. Granek and M.E. Cates, Phys. Rev. A{\bf 46},
3319 (1992).

\bibitem{VoBiKa} C.G. Vonk, J.F. Billman, and E.W. Kaler, J. Chem.
Phys. {\bf 88}, 3970 (1988).

\bibitem{Siggia} E.D. Siggia, Phys. Rev. A{\bf 20}, 595 (1979).

\bibitem{LifSlo} I.M. Lifshitz and V.V. Sloyzov, J. Chem. Phys. Solids
{\bf 19}, 35 (1961).

\bibitem{Landau} L.D. Landau and E.M. Lifshitz, {\it Fluid Mechanics}
(Pergamon Press, Oxford, 1959).

\bibitem{GolLub} L. Golubovi\'c and T.C. Lubensky, Phys. Rev. B{\bf 39},
12110 (1989).

\bibitem{KadMartin} L.P. Kadanoff and P.C. Martin, Ann. Phys. {\bf
24}, 419 (1963).

\bibitem{PierMarc} P. Pieruschka and S. Mar\u celja, J. Phys. II
France {\bf 2}, 235 (1992).

\bibitem{GoGoos94} G. Gompper and J. Goos, Phys. Rev. E (to appear).

\end{thebibliography}
\end{document}